\documentstyle[12pt]{article}
           \newcommand{\G}{\Gamma}
\renewcommand{\d}{\delta}         

\newcommand{\eps}{\epsilon}

           \renewcommand{\S}{\Sigma}

\newcommand{\vf}{{\varphi}}

\newcommand{\cx}{{X}}
\newcommand{\cxz}{{X_0}}
\newcommand{\cxo}{{X_1}}
\newcommand{\cxd}{{X_2}}
\newcommand{\cxt}{{X_3}}

\newcommand{\cq}{Q}
\newcommand{\cqi}{Q_i}
\newcommand{\cqz}{Q_0}
\newcommand{\cqo}{Q_1}
\newcommand{\cqd}{Q_2}
\newcommand{\cqt}{Q_3}

\def\limar#1#2{\,\raise0.3ex\hbox{$\longrightarrow$\kern-1.5em\raise-1.1ex
\hbox{$\scriptstyle{#1\rightarrow #2}$}}\,}
\def\limarr#1#2{\,\raise0.3ex\hbox{$\longrightarrow$\kern-1.5em\raise-1.3ex
\hbox{$\scriptstyle{#1\rightarrow #2}$}}\,}
\def\limlar#1#2{\ \raise0.3ex
\hbox{$-\hspace{-0.5em}-\hspace{-0.5em}-\hspace{-0.5em}
\longrightarrow$\kern-2.7em\raise-1.1ex
\hbox{$\scriptstyle{#1\rightarrow #2}$}}\ \ }

\newcommand{\Ibb}[1]{ {\rm I\ifmmode\mkern
            -3.6mu\else\kern -.2em\fi#1}}
\newcommand{\ibb}[1]{\leavevmode\hbox{\kern.3em\vrule
     height 1.2ex depth -.3ex width .2pt\kern-.3em\rm#1}}

\newcommand{\rational}{{\kern .1em {\raise .47ex
\hbox{$\scripscriptstyle |$}}
    \kern -.35em {\rm Q}}}

\newcommand{\intsz}{{\int\limits_{s_0}^{\infty}}}

\newcommand{\pad}[2]{{\frac{\partial #1}{\partial #2}}}
\newcommand{\padd}[2]{{\frac{\partial^2 #1}{\partial {#2}^2}}}

\def\<{\langle}
\def\>{\rangle}

\newcommand{\be}{\begin{equation}}
\newcommand{\ee}{\end{equation}}
\newcommand{\bn}{\begin{eqnarray}}
\newcommand{\en}{\end{eqnarray}}
\newcommand{\bnn}{\begin{eqnarray*}}
\newcommand{\enn}{\end{eqnarray*}}
\newcommand{\ba}{\begin{array}}
\newcommand{\ea}{\end{array}}

\newcommand{\biz}{\begin{itemize}}
\newcommand{\eiz}{\end{itemize}}
\newcommand{\ben}{\begin{enumerate}}
\newcommand{\een}{\end{enumerate}}
\newcommand{\cost}{$\cos\,\vartheta$ }
\newcommand{\sts}{\sigma _{tot}\,(s) }

\newcommand{\ses}{\sigma _{el}\,(s) }
\newcommand{\tses}{\tilde \sigma _{el}\,(s) }

\newcommand{\am}{amplitude }

\newcommand{\any}{analyticity }
\newcommand{\anyp}{analyticity. }
\newcommand{\anl}{analytical }

\newcommand{\con}{condition }
\newcommand{\com}{commutative }
\newcommand{\coc}{commutative case }
\newcommand{\coy}{commutativity }

\newcommand{\drs}{dispersion relations }
\newcommand{\drsp}{dispersion relations. }

\newcommand{\es}{elastic scattering }
\newcommand{\esa}{elastic scattering amplitude }
\newcommand{\esas}{elastic scattering amplitudes }

\def\fmb{Froissart-Martin bound }
\def\fmbp{Froissart-Martin bound. }
\newcommand{\fu}{function }

\def\jld{Jost-Lehmann-Dyson  }

\def\lee{Lehmann ellipse }
\def\leep{Lehmann ellipse. }
\def\lsz{Lehmann-Symanzik-Zimmermann (LSZ) }

\def\mae{Martin ellipse }
\def\maep{Martin ellipse. }

\newcommand{\nc}{noncommutative }
\newcommand{\ncy}{noncommutativity }

\newcommand{\qft}{quantum field theory }

\def\rep{representation }
\def\repp{representation. }

\newcommand{\sca}{scattering }
\newcommand{\sam}{scattering amplitude }
\newcommand{\samp}{scattering amplitude. }
\newcommand{\sams}{scattering amplitudes }
\newcommand{\spsp}{space-space }

\newcommand{\lsr}{\ln^2\,\frac s {s_0}}

\newcommand{\sz}{s_0 }

\newcommand{\al}{\alpha}

\newcommand{\de}{\delta }
\newcommand{\decxs}{\delta\,(X^2) }

\newcommand{\lao}{{\lambda}_1 }
\newcommand{\lad}{{\lambda}_2 }

\newcommand{\txz}{\tau\,(x_0)}

\def\thmn{\theta_{\mu\nu}}

\def\thod{\theta_{12}}
\def\thdo{\theta_{21}}

\def\ss{$\theta_{0i}=0$}

\newcommand{\jox}{j_1\,(x)}
\newcommand{\jdx}{j_2\,(x)}
\newcommand{\jix}{j_i\,(x)}

\newcommand{\joz}{j_1\,(0)}
\newcommand{\jdz}{j_2\,(0)}

\newcommand{\joxh}{j_1\,\left({\frac x 2}\right)}

\newcommand{\jdmxh}{j_2\,\left({- \frac x 2}\right)}

\newcommand{\kjoxjdy}{[j_1\,(x), j_2\,(y)]}
\newcommand{\kjohxjdhy}{[j_1\,(\hat x), j_2\,(\hat y)]}

\newcommand{\kjoxhjdmxh}{\left[j_1\,\left({\frac x 2}\right),
j_2\,\left(- {\frac x 2}\right)\right]}

\newcommand{\kz}{k_0}

\newcommand{\pz}{p_0}
\newcommand{\po}{p_1}
\newcommand{\pzp}{p'_0}

\newcommand{\xz}{x_0}

\newcommand{\xt}{x_3}

\newcommand{\yo}{y_1}
\newcommand{\yd}{y_2}

\newcommand{\qz}{q_0}
\newcommand{\qo}{q_1}
\newcommand{\qd}{q_2}
\newcommand{\qt}{q_3}
\newcommand{\qzp}{q'_0}
\newcommand{\tq}{\tilde q}

\newcommand{\uz}{u_0}

\newcommand{\uzp}{u_0'}

\newcommand{\xmys}{{(x - y)}^2}
\newcommand{\xzmyzs}{{(x_0 - y_0)}^2}
\newcommand{\xtmyts}{{(x_3 - y_3)}^2}

\newcommand{\dx}{d\,x}
\newcommand{\dxz}{d\,x_0}
\newcommand{\dxo}{d\,x_1}
\newcommand{\dxd}{d\,x_2}
\newcommand{\dxt}{d\,x_3}

\newcommand{\fcx}{F\,(X) }
\newcommand{\fpk}{F\,(p', k',p, k)}

\newcommand{\fsct}{F\,(s, \cos \, \vartheta) }
\newcommand{\casct}{A\,(s, \cos \, \vartheta)}
\newcommand{\cast}{A\,(s, t)}
\newcommand{\casz}{A\,(s, 0)}
\newcommand{\castz}{A\,(s, t_0)}

\newcommand{\plct}{P_l\,(\cos \, \vartheta)}
\newcommand{\pl}{P_l\,(x)}

\newcommand{\plppo}{P_{l + 1}^{'}\,(x)}
\newcommand{\plpmo}{P_{l - 1}^{'}\,(x)}

\newcommand{\fcq}{F\,(Q) }

\newcommand{\ftq}{F\,(\tilde q) }
\newcommand{\fcqp}{F\,(Q') }

\newcommand{\fsx}{f\,(x)}
\newcommand{\fsrx}{f^r\,(x)}

\newcommand{\fsrxzxt}{f^r\,(x_0, x_3)}

\newcommand{\fsq}{f\,(q)}
\newcommand{\fsoq}{f_1\,(q)}
\newcommand{\fsdq}{f_2\,(q)}
\newcommand{\fsrq}{f^r\,(q)}

\newcommand{\tals}{\tilde {a_l}\,(s)}

\newcommand{\tfls}{\tilde {f_l}\,(s)}

\begin{document}
\begin{titlepage}
\renewcommand{\baselinestretch}{2.0}
\begin{center}

{\Large{\bf{ Jost-Lehmann-Dyson Representation, Analyticity in Angle 
Variable and Upper Bounds in Noncommutative Quantum Field Theory }}}
\vskip .7cm

{\bf{\large{ M. N. Mnatsakanova$^{a}$
 \ \ and \ \ Yu. S. Vernov$^{b}$}}}

{\it $^a$Institute of Nuclear Physics, Moscow State University,\\
Moscow, 119992 Russia\\

$^{b}$Institute for Nuclear Research,
Russian Academy of Sciences,\\Moscow, 117312 Russia }

\setcounter{footnote}{0}

{\bf Abstract}
\end{center}

The existence of Jost-Lehmann-Dyson representation analogue has been 
proved in framework of space-space noncommutative quantum field theory. On 
the basis of this representation it has been found that some class of 
elastic amplitudes admits an analytical continuation into complex $\cos 
\vartheta$ plane and corresponding domain of analyticity is Martin 
ellipse. This analyticity combined with unitarity leads to 
Froissart-Martin upper bound on total cross section.

\end{titlepage}
\newpage
\setcounter{page}{1}

\section{Introduction}

The proof of \anl properties of \esas in $\cos\,\vartheta$, ($\vartheta$ is a \sca
angle) is one of the most important achievements of \qft (QFT). The first
step was done by Lehmann \cite{Leh}, who proved that $\pi-N$ \esa is an
\anl \fu of $\cos\,\vartheta$ in some ellipse (Lehmann ellipse). Martin
\cite{Mar66} derived that using \any with respect to energy
variable it is possible to enlarge sufficiently the above mentioned
domain of \anyp The exact size of this domain was established in \cite
{Som, Bes}. This domain is named Martin ellipse.

The \any in \cost plane together with unitarity leads to the very
important bounds on high energy behaviour of \samp The first rigorous
bound, which follows from \any in \lee was obtained in the work of
Greenberg and Low \cite {GrL}.

In accordance with this bound at $s \to \infty$
\be\label{st}
\sts \leq C\,s\,\lsr ,
\ee
$\sts$ is a total cross-section, $C$ and $\sz$ are some constants. $s$ is
the usual variable, in this paper we deal only with a center of mass 
system.

Froissart \cite{Fro} showed that the bound considerably stronger then 
(\ref{st}) follows from double dispersion relations, namely 
\be\label{stf}
\sts \leq C\,\lsr .
\ee
Martin \cite{Mar63} first demonstrated that this bound is valid under
much weaker  \con on the domain of \any in \cost plane and then proved
that the necessary domain of \any really exists in the framework of
axiomatic QFT \cite{Mar66}. The bound (\ref{stf}) is named \fmbp The
final step in derivation of the best axiomatic upper bound on total
cross-section was done in \cite{LuM}, where it was proved that
\be\label{stlm}
\sts \leq \frac{\pi}{m^2_{\pi}}\,\lsr,
\ee
$m_{\pi}$ is $\pi$-meson mass.

Precisely, slightly stronger bound can be obtained. Namely, at $s \to
\infty$ (see, e.g. reviews \cite{Roy} and \cite{VM})
\be\label{stlms}
\sts \leq \frac{\pi}{m^2_{\pi}}\,\ln^2\,\frac s {{(\ln\,s)}^{\frac 3 2}}.
\ee
The rigorous upper bounds were also found for a forward differential
cross-section. The best bound of this kind was obtained by Singh and Roy
\cite{SR}:
\be\label{sir}
|F\,(s)| \leq
\frac{s}{8\,m_{\pi}\,\sqrt{\pi}}\,
{\sqrt{\sts }}\,\ln{\frac{s}{\sts}}.
\ee
The bounds for non-forward scattering were also obtained (see reviews
\cite{Roy, LMKh}).

At present \nc quantum field theory (NC QFT) is regarded as one of the
most attractive possibilities to consider interaction at very short
distances and so at very high energies. The study of such theories
acquired an additional interest after it was shown that they appear
naturally, in some cases, as low-energy effective limits of open string
theory \cite{SW}. In \nc \qft the coordinate operators 
satisfy the commutation relations
\be\label{cr}
[\hat{x}_\mu,\hat{x}_\nu]=i\theta_{\mu\nu},
\ee
where
$\theta_{\mu\nu}$ is a constant antisymmetric matrix of dimension
(length)$^2$.

The implications of the modern ideas of noncommutative geometry
\cite{Connes} in physics have been lately of great interest, though
attempts can be traced back as far as 1947 \cite{Snyder}. Plausible new
arguments for studying NC QFT have been offered in \cite{SW, Dopli} (for a
review, see \cite{DN}).  Thus it is very important to investigate \anl
properties of \sams in NC QFT.

We shall consider throughout this paper only the case of
space-space noncommutativity, i.e. $\theta_{0i}=0$, since theories
with space-time \ncy can be obtained as low-energy effective limits from 
string theory only in special cases \cite{Go2}. Besides, there are 
problems with unitarity \cite{GM} and causality \cite{SSB,cnt} in the 
general case.

In our papers \cite{CMTV, VM04} it was shown that in space-space \nc field
theory a forward \esa has the same \anl properties as in commutative one.
Let us point out that the first step in the derivation of similar \anl
properties has been done in \cite{Liao}.

In \com QFT the proof of \anl properties of \esas is based on the local
\coy (microcausality) condition, that is
\be\label{lcc}
\kjoxjdy = 0, \quad \mbox{if} \quad \xmys < 0,
\ee
where $\jix, \; i = 1,2$, is an interacting current (see \cite{StW}, eq.
(3-34)). Below we take $\jox$ to mean a nucleon current and  $\jdx$ -
a $\pi$-meson one.

In \spsp NC field theory, that is \ss, we can, without lost of generality,
consider the case, when only $\thod = - \thdo \neq 0$ \cite{AlG}, in other
words $\xz$ and $\xt$ are commutative variables and one can assume
\cite{AlG} that condition (\ref{lcc}) has the \nc analogue:
\be\label{sscc}
\kjohxjdhy = 0, \quad \mbox{if} \quad \xzmyzs - \xtmyts < 0. 
\ee
In our papers \cite{CMTV, VM04} it was proved that condition
(\ref{sscc}) is sufficient to derive the usual forward \drsp This fact
would play a crucial role in obtaining the final domain of \any in angle
plane in \nc case. We show that this domain is the same with the
largest rigorously proved domain of such an \any in \com case -  Martin
ellipse. This fact gives us the possibility to prove \fmb in case
of \spsp NC QFT. As we consider only this variant of \nc theory we do not
mention below this point. The derivation of  \jld \rep in NC \spsp QFT was done also
in the paper \cite{ChT}, which appeared simultaneously with the first version of this work. In \cite{ChT} Froissart-Martin bound has been obtained, the result is based on the assumptions different from ours.    

Our paper is organized as follows. We consider \esa of two spinless
particles with masses $m$ (meson) and $M$ (nucleon). We believe that, as
well as in commutative case, obtained results would coincide with the
results for $\pi-N$-scattering after averaging over spin.
Our results can be also extended on other processes.

First we obtain the analogue of \jld \repp Using this \rep we derive
the \any of the \am in question in \leep Following the way proposed in
\cite{Mar66} - \cite{Bes} we can extend this ellipse up to \maep
In this extension as well as in \coc is the existence of dispersion
relations plays a crucial role.

In Appendix for the reader's convenience we give derivation of \fmb 
in its strongest form under the weakest assumptions.

Let us point out that in \nc case \fmb can be stronger than in the usual case
(see (\ref{stlms})).  The matter is that really \fmb contains the factor
$\frac{\ses}{\sts}$, where $\ses$ is an elastic scattering
cross-section. In \nc case \fmb contains
the analogous factor $\frac{\tilde \ses}{\sts}$, where $\tilde \ses$ is an 
elastic cross-section in the case when momentums of initial particles are 
orthogonal to \nc plane.

\section{\jld Representation }

Let us consider the matrix element
\be\label{mu}
f\,(x) = \langle p'\,|\,\kjoxhjdmxh\,|\,p \rangle,
\ee
where $|\,p \rangle$ and $|\,p' \rangle$  are arbitrary states with
momentum $p$ and $p'$ correspondingly. To simplify notations we omit
$\hat{}$ above $x$. Owing to condition (\ref{sscc})
\be\label{8}
\fsx = 0, \quad \mbox{if} \quad  {\xz}^2 - {\xt}^2 < 0.
\ee
Fourier transformation of this matrix element is:
\be\label{gft} f\,(q) =
\int\,e^{i\,q\,x}\,f\,(x)\,\dx.
\ee
We omit unessential numerical factor.
Carry out the integration in this
expression over \nc variables $x_1$ and $x_2$ we obtain
\be\label{anyf}
f\,(q) \equiv f\,(q_0, q_3) =  \int\,e^{i\,(q_0\,x_0 -
q_3\,x_3)}\,f\,(x_0, x_3)\,\dxz\,\dxt.
\ee
Here
\bnn
f\,(x_0, x_3) \equiv \int\,f\,(x)\,e^{- i\,(q_1\,x_1 -
q_2\,x_2)}\,\dxo\,\dxd.
\enn
We do not write down the dependence $f\,(x_0, x_3)$ on $q_1$ and $q_2$.
Here and in what follows (except as otherwise noted) $q$ is a two
dimensional vector $q = (\qz, \qt), \; q^2 = {\qz}^2 - {\qt}^2$.  The
corresponding Fourier transformation is:
\be\label{frfsq}
\fsq \equiv
\int\,e^{i\,q\,x}\,\fsx\,\dxz\,\dxt, \quad f\,(x) \equiv f\,(x_0, x_3).
\ee
To use efficiently the condition (\ref{8}) let us (similar to the \com
case) associate \fu $\fsx$ with the \fu $\fcx$ in four-dimensional
space, where
\be\label{9}
\fcx  =  4\,\pi\,\fsx\,\decxs,
\ee
\bnn
\cxz  =  \xz, \: \cxo  =  \xt, \: \cxd  =  \yo,\:
\cxt = \yd,  \quad  {\cx}^2  =  {\xz}^2 - {\xt}^2 - {\yo}^2 -
{\yd}^2.
\enn
In accordance with \con (\ref{8})
\be\label{10}
\fsx  =
\frac {1} {4\,{\pi}^2}\,\int\,\fcx\,d^2\,y.
\ee
Indeed, by definition (\ref{9})
\be\label{11}
\fsx  =  \frac {1} {4\,\pi}\,\int\,\fcx\,d\,y^2 =
\left \{\begin{array}{lcl}
\fsx & \mbox{at} & x^2 \geq 0, \\
0 & \mbox{at} & x^2 < 0
\end{array}\right. .
\ee
As in our case $\fsx = 0$ if $x^2 < 0$, we checked eq. (\ref{10}).

Let us consider four-dimensional momentum space $\cqi$: $\cqz = \qz$,
$\cqo = \qt$, $\cqd = \lao$, $\cqt = \lad$; ${\cq}^2 = {\qz}^2 - {\qt}^2 -
{\lao}^2 - {\lad}^2$.
\
It is easy to see that
\be\label{12}
{\Box}_4\,\fcq = 0, \quad {\Box}_4 \equiv \padd {}{\qz} - \padd {}{\qt} -
\padd {}{\lao} - \padd {}{\lad},
\ee
where
\bnn
\fcq = \int\,e^{i\,Q\,X}\,\fcx\,d^4\,X.
\enn
The correspondence between $\fsq$ and $\fcq$ can be easily obtained:
\bnn
\fsq = \ftq,\quad \mbox{where} \quad \tq = (\qz, \qt, 0, 0).
\enn

The following consideration is similar to the one in the usual
(commutative) case with the only difference that now we have
four-dimensional space instead of six-dimensional one. The general
consideration in space of arbitrary dimensions was done in Vladimirov's
book \cite{Vlad}, see similar consideration in four-dimensional case in
\cite{BoSh, Schw}.

In accordance with eq. (\ref{12}) we can represent $\fcq$ in the form:
\be\label{13}
\fcq = \int\limits_{\S}\,d\,{\S}'\left[D\,(Q - Q')\,\pad
{\fcqp}{{\G}'} + \fcqp\,\pad {D\,(Q - Q')}{{\G}'}\right],
\ee
where $\S$ is some three-dimensional hypersurface, $\pad {}{{\G}'}$ is a
conormal derivation on it and \fu $D\,(Q)$ satisfies eq. (\ref{12}) as
well as the following condition:
\ben
\item
$$ D\,(Q) \to 0 \quad \mbox{if} \quad \cqz \to 0;
$$
\item
$$
\pad {D\,(Q)}{\cqz} \to \de\,(\vec{Q}) \quad \mbox{if} \quad \cqz \to 0;
\quad \de\,(\vec{Q}) = \de\,(\cqo)\,\de\,(\cqd)\,\de\,(\cqt);
$$
\item
$$
\padd {D\,(Q)}{\cqz} \to 0 \quad \mbox{if} \quad  \cqz \to 0.
$$
\een
The necessary \fu would be
\be\label{14}
D\,(Q) =
\frac{-i}{{(2\,\pi)}^3}\int\,\eps\,(\xz)\,e^{i\,Q\,X}\,\de\,(X^2)\,d^4\,X.
\ee
(The proof one can find in \cite{Vlad}).

We can, as well as in \com case (see \cite{BoSh}, Ch. 10), simplify eq.
(\ref{13}) integrating the second term by parts.
Thus we can rewrite expression (\ref{13}) as follows:
\be\label{15}
\fcq = \int\,d^4\,Q'\,D\,(Q - Q')\,\psi\,(Q').
\ee
At the moment it is convenient to consider
integral in (\ref{14}) formally as integral over all space (really
$\psi\,(Q')$ contains the term $\de\,(\S)$).
Below we see that integration limits are really determined by the
properties of $\fsq$, which follow from translation invariance and
spectral condition.
Using the explicit expression for $D\,(Q)$ (see \cite{Vlad}):
\be\label{16}
D\,(Q) = \frac{1}{2\,\pi}\,\eps\,(\cqz)\,{\de}\,(Q^2)
\ee
we come to the final expression:
\be\label{17}
F\,(Q) =
\int\,d^4\,Q'\,\eps\,(Q_0 - Q'_0)\,\de\,{(Q - Q')}^2\,\psi\,(Q').
\ee
From eq. (\ref{17}) it follows directly that $\fsq = \ftq$ satisfies the
representation:
\be\label{18}
\fsq = \int\,d^4\,u'\,\eps\,(\qz - \uzp)\,{\de}\,{(q - u')}^2\,\vf\,(u') 
\ee
(we change the notations: $Q' = u', \; \psi\,(Q') = \vf\, (u')$).

Let us proceed to the similar expression for
\be\label{19}
\fsrq =
\int\,d\,\qz\,d\,\qt\,\txz\,e^{i\,(\qz\,\xz - \qt\,\xt)}\,\fsx,
\ee
where
$$
\txz = 1, \quad \xz \geq 0; \qquad \txz = 0, \quad \xz < 0.
$$
It is easy to show that
\be\label{20}
\fsrq = \frac{i}{2\,\pi}
\int\,d\,\qzp\,\frac{f\,(\qzp, \vec{q})}{\qzp - \qz}, \quad Im\,\qz >0.
\ee
Using expressions (\ref{19}) and (\ref{18}) we can easily make necessary
integrations in (\ref{20}) and finally obtain:
\be\label{21}
\fsrq = \frac{i}{2\,\pi}
\int\,d^4\,u'\,\frac{\vf\,(u')}{{(q - u')}^2}, \quad Im\,\qz >0.
\ee

We represent space vector $\vec{u'}$ as the linear combination of two 
orthogonal vectors: $\vec{u}$, belonging to the plane, which contains the 
axis $\lao$ and $\vec{q}$, and $\lad\,\vec{e}$, where $\vec{e}$ is a  
unit vector, directed along the axis $\lad, \; u'_0 \equiv 
u_0$.  Eq. (\ref{21}) can be written as follows
\be\label{26'}
\fsrq = \frac{i}{2\,\pi}
\int\,d^3\,u\,\int\,d\,{\lad}^2\frac{\vf\,(u, {\lad}^2)}
{{(q - u)}^2 - {\lad}^2}, \quad  Im\,\qz >0.  
\ee

Now we find the restrictions on the domain of integration in eqs. 
(\ref{18}) and (\ref{26'}). To this end we use the spectral properties of 
$\fsq$. First let us represent $\fsq$ in eq. (\ref{gft}) as $\fsq = \fsoq
-\fsdq$, where
\be\label{21a_1}
\fsoq = \int\,e^{i\,q\,x}
\,\langle p'\,|\,\joxh\,\jdmxh\,|\,p \rangle\,\dx,
\ee
\be\label{21a_2}
\fsdq = \int\,e^{i\,q\,x}
\,\langle p'\,|\,\jdmxh\,\joxh\,|\,p \rangle\,\dx.
\ee
Surely here $q$ is a four dimensional vector.

Taking into account that translation invariance survives in NC QFT and
using it as well as completeness of basic vectors $|P_n\rangle$,
($P_n$ is the momentum of the $n$  state) we obtain in Breit system, that
is $\vec{p} + \vec{p'} = 0$:
\be\label{21b_1}
\fsoq = \sum\limits_n\,\langle p'
\,|\,\joz\,|\,P_n\, \rangle \,\langle
\,P_n\,|\,\jdz\,|\,p \rangle\,
\de\,(q_0 + a - \sqrt{M_n^2 + {\vec{q}}^2})
\ee
\be\label{21b_2}
\fsdq = \sum\limits_n \,\langle p'\,|\,\jdz\,|\,P_n\,
\rangle \,\langle\,P_n\,|\,\joz\,|\,p \rangle\,\de\,(- q_0
+ a - \sqrt{M_n^2 + {\vec{q}}^2}),
\ee
where $a = \frac{\pz + \pzp}{2}$,
$M_n$ are masses of intermediate states.

Thus $\fsq = 0$ if the following double inequality is satisfied:
\be\label{31}
a -  \sqrt{{\vec{q}}^2 + m_2^2} < \qz < \sqrt{{\vec{q}}^2 + m_1^2} - a,
\ee
where $m_1$ and $m_2$ are the minimal masses of intermediate states
$|\,P_n\,\rangle$. Just the same conditions are valid in commutative case
\cite{BoSh}.

Let us consider the case when $\qo = \qd = 0$. In this case we can use the
same notation both for four-dimensional vector $(\qz, \qt, 0, 0)$ and
two-dimensional one $(\qz, \qt)$.

Condition (\ref{31}) can be written in the form:
\be\label{32}
S_-\,(\vec{q}) < \qz < S_+\,(\vec{q}), \quad \vec{q} = \qt.
\ee
We determine  two surfaces $\sigma_{\pm}$, such that $\qz = 
S_{\pm}\,(\vec{q})$.

In order to the condition (\ref{32}) be satisfied automatically 
\cite{BoSh, Schw} we choose as usual the domain of integration in 
(\ref{18}) so that $\de$-\fu in (\ref{18}) be zero, when $\qz$ satisfies 
the double inequality (\ref{32}). 

As it follows from eq. (\ref{18}) $\fsq \neq 0$ only if 
\be\label{33} 
{(q - u)}^2 = {\lad}^2.  
\ee 
Eq. (\ref{33}) determines 
two-branch hyperboloid. Following Dyson let us call this hyperboloid 
admissible if its upper branch has no points below $\sigma_+$  and its 
lower one has no points upper $\sigma_-$, that is:  
$$ 
u_0 + \sqrt{{(\vec{q} - \vec{u})}^2 + {\lad}^2} \geq  S_+\,(\vec{q}), 
$$ 
$$ 
u_0 - \sqrt{{(\vec{q} - \vec{u})}^2 + {\lad}^2} \leq  S_-\,(\vec{q}).  
$$ 
These conditions would be satisfied for any $\vec{q}$ if
$$
u_0 \geq \max_q{(S_+\,(\vec{q}) - \sqrt{{(\vec{q} - \vec{u})}^2 +
{\lad}^2})},
$$
$$
u_0 \leq \min_q{(S_-\,(\vec{q}) + \sqrt{{(\vec{q} - \vec{u})}^2 +
{\lad}^2})}.
$$
The necessary calculations are similar to ones drawn in \cite{BoSh, Schw}.
As a result we have:
\be\label{34}
|\uz| + |\vec{u}| \leq a,
\ee
\be\label{35}
{\lad}^2 \geq max \{ {(m_2 - \sqrt{{(a - \uz)}^2 - {\vec{u}}^2})}^2,
{(m_1 - \sqrt{{(a + \uz)}^2 - {\vec{u}}^2})}^2\}.
\ee

\section{ Elastic Scattering Amplitude and  Analyticity \\ in Lehmann 
Ellipse }

Let us consider $\pi-N$ \esa for the process, in which $\pi$-meson with
mass $m$ has the initial momentum $k$ and final $k'$ and nucleon with
mass $M$ has the initial momentum $p$ and final $p'$. As it was shown in
\cite{CMTV} the usual \lsz formulas are valid in \nc \spsp field theory.
Thus up to the numerical factors and the term, which is a polynomial in
energy, we can write
\be\label{afr}
\fpk = \int\,e^{\frac{i\,(p - k)\,x}{2}}\fsrx\,\dx,
\ee
where
\bnn
\fsrx = \tau\,(x_0) \,\langle p', k'\,|\,\kjoxhjdmxh\,|\,0 \rangle,
\enn
$\jox$ is a nucleon current and $\jdx$ is a $\pi$-meson current.

Let us consider \sam in the center of mass system.
To use \jld \rep we represent \sam in the form:
\be\label{37}
\fpk = \int\,e^{i\,\left[{\frac{p_0 -k_0}{2}}\,x_0 -
p_3\,x_3\right]}\,\fsrxzxt\,\dxz\,\dxt,
\ee
where
\bnn
\fsrxzxt =
\int\,e^{- i\,(p_1\,x_1 + p_2\,x_2)}\,\fsrx\,\dxo\,\dxd.
\enn
As before we consider the case $p_1 = p_2 = 0$.

To derive analyticity in Lehmann ellipse let us use the representation
(\ref{26'}) for \esa
\be\label{38}
\fpk = \frac{i}{2\,\pi}
\int\,d^3\,u\,\int\limits_{\min\,{\lad}^2}^{\infty}\,d\,{\lad}^2\,
\frac{\vf\,(u,{\lad}^2, p', k', \thmn)}
{{\left(\uz - \frac {\pz -\kz}{2}\right)}^2 - {(\vec{u} - \vec{p})}^2 - 
{\lad}^2}.  
\ee

We choose axis $\lao$ so that it would belong to the plane formed by 
vectors $\vec{q}$ and $\vec{p'}$. Let us denote the angle between 
$\vec{u}$ and $\vec{p'}$ as $\alpha$, then $\vec{u}\,\vec{p} = 
|\vec{u}|\,|\vec{p}|\,\cos\,(\vartheta - \al)$, where $\vartheta$ is the 
angle between $\vec{p}$ and $\vec{p'}$.
Eq. (\ref{38}) is similar to
\com one \cite{LMKh} with the only and evident difference that in our 
case we have no additional integrations.  Let us stress that in \nc case 
numerator has an additional dependence of $\thmn$, but similarly to the 
usual case all dependence of $\cos \vartheta$ is contained in denominator.

Now let us consider the dominator in the last integral in (\ref{38}). 
The direct calculation shows that denominator is
$$
- 2\,|\vec{u}|\,|\vec{p}|\,\sin\,\beta \,[y -\cos\,(\vartheta - \al)], 
$$ 
where 
$$ 
y = \frac{{\vec{u}}^2 + {\vec{p}}^2 + {\lad}^2 - {\left(u_0 - \frac {p_0 - 
k_0}{2}\right)}^2}{2\,|\vec{u}|\,|\vec{p}|\,\sin\,\beta}.  
$$ 
Now our goal is to calculate the minimal value of $y \equiv y_0\,(s)$. 
It is significant that calculations in \nc case reduce to the similar one 
in \com case when  $\po = p_2 =0$.  In other words, we 
consider \es process in which $\vec{p}$ is orthogonal to \nc plane. 
As forward \sam is a function of $s$ only, it does not depend on this 
additional condition.

As above we use translation invariance and spectral condition in order
to obtain the necessary constrains on $y_0\,(s)$. Actually, repeating the
steps we have made in derivation of double inequality (\ref{31}) we 
obtain the similar one, but now $a = {({\pz}' + {\kz})'}/2$. Evidently the 
constrains (\ref{34}) and (\ref{35}) are valid. To estimate minimum of $y$ 
we can use directly the results obtained in \com case.  
Indeed, the restrictions (\ref{34}), (\ref{35}) contain only 
${\vec{u}}^2$. 
This minimum is searched among all possible values of $\vec{u}$, which 
include the case when $\vec{u}$ belongs the plane containing 
vectors $\vec{p}$ and $\vec{p'}$, that actually is our case. 
As in commutative case the analogous minimum is realized just in case when 
this additional condition  is satisfied \cite{Leh}, the results coincide 
both in commutative case and noncommutative one.
The derivation of this minimum also uses the masses 
of the lowest intermediate states, which are evidently the same with \com 
case.  As this minimum is more than one the right-hand side of eq. 
(\ref{38}) is well-defined also for nonphysical $\cos \vartheta$, 
resulting in analyticity of the amplitude under consideration with respect 
to $\cos \vartheta$ in the domain, which is Lehmann ellipse. Let us recall 
that \lee is the ellipse with focuses in points $\pm 1$ and with major 
half-axis $y_0\,(s)$.  The exact value of $y_0\,(s)$ one can find in 
\cite{LMKh}, here we only mentioned that at $s \to \infty$ 
$$ 
y_0\,(s) 
\cong 1 + \frac{c\,(m, M)}{s^2} \quad \mbox{i.e.} \quad t_0\,(s) \cong 
\frac{2\,c\,(m, M)}{s}.  
$$ 
Let us remind that at high energies transfer 
momentum is represented by scattering angle as follows:  $\cos \vartheta = 
1 + {2\,t}/s$.

Using unitarity and spectral properties of scattering amplitude it is easy
to show that as in \com case imaginary part of \esa - $\casct$ is an
analytical function in the domain, which is larger than analogous
domain for $\fsct$ - large \lee \cite{Leh}, see also \cite{LMKh}.

\section{ Unitary Constrains on Particle Amplitudes }

For the elastic process, when $p_1 = p_2 = 0$, we can say that scattering
amplitude is a function of $s$ and $\cos \vartheta$ (the dependence of this
function on $\thmn$ does not change the results derived below.
Thus for the elastic scattering amplitude in question we can write
$$
F\,(p', k', p, k) = F\,(s, \cos \, \vartheta, \thmn)
\equiv  F\,(s, \cos \, \vartheta).
$$
We can expand $F\,(s, \cos \, \vartheta)$ in the Legendre polynomials:
\be\label{39}
F\,(s, \cos \, \vartheta) = 2\,\sum\limits_0^{\infty}\, (2\,l + 1)\,\tilde
{f_l}\,(s) P_l\,(\cos \, \vartheta)
\ee
as $F\,(s, \cos \, \vartheta)$ is an analytical function in Lehmann
ellipse \cite{Whi}.

As we further deal with       
$\fsct$ at $s \to \infty$, we substitute the factor in front of the sum by
its asymptotic value.

Surely $\tilde {f_l}\,(s)$ depends on $\thmn$,
but this dependence does not change constrains, which we obtain below.  We
use the notation $\tilde {f_l}\,(s)$ instead of the usual  $f_l\,(s)$ in order
to emphasize that we consider only the specific class of elastic
scattering amplitudes. Owing to orthogonality of the Legendre polynomials
\be\label{40}
\tilde {f_l}\,(s) = \int\limits_{- 1}^1\, F\,(s, \cos\,\vartheta)
\,P_l\,(\cos \,\vartheta)\,d\,\vartheta.
\ee
Let us point out that for forward scattering eq. (\ref{39}) is the general
expression as $F\,(s, 1)$ does not depend on the direction of $\vec{q}$.
Now let us consider the special differential cross section and thus
special elastic scattering with $\tilde {\sigma_{el}}\,(s)$,
corresponding to the chosen class of amplitudes.  That is
\be\label{41}
s\,\tilde {\sigma_{el}}\,(s) = \int\limits_{- 1}^1 F\,(s, \cos
\, \vartheta) F^*\,(s, \cos \, \vartheta)\,d\,\cos \, \vartheta.
\ee
Using the eq. (\ref{39}), we obtain
\be\label{42}
s\,\tilde
{\sigma_{el}}\,(s) = \sum\limits_0^{\infty} \, (2\, l + 1)\,{|\tilde
{f_l}\,(s)|}^2
\ee
In accordance with the optical theorem, which follows
only from the unitarity and thus has to be also valid in NC case we have:
\be\label{43}
s\,\sigma_{tot}\,(s) = \sum\limits_0^{\infty} \,(2\, l + 1)\,\tilde
{a_l}\, (s), \quad  \tals \equiv Im\,\tfls,
\ee
we use here the convenient normalization for $\sts$ and $\ses$.

As
$$
\tilde {\sigma_{el}}\,(s) <  \sigma_{tot}\,(s)
$$
we obtain that
\be\label{44}
\tilde {a_l}\,(s) \geq {|\tilde {f_l}\,(s)|}^2,
\ee
which is the usual unitarity constrain on partial amplitudes. Thus we
have no problem with the derivation of Froissart-Martin bound. 
Let us point out that Froissart-Martin bound in NC case can be stronger 
than in the usual one as we deal only with $\tilde {\sigma_{el}}\,(s)$ 
(see the end of Introduction).

\section{ Martin Ellipse and Froissart-Martin Bound in NC QFT }

The possibility to enlarge \lee up to the Martin's one is caused on the one
hand by unitary constrains on partial amplitudes and on the other hand
by \drs (DR).
Let us remind that \mae has focuses at the same point as the Lehmann's 
one, its extremely right point is $t_0\,(s) = 4\,m_{\pi}^2$.

Strictly speaking the validity of DR is necessary not
only for forward direction, but also for arbitrary small negative $t$,
which is the usual variable. In \nc case DR were proved only for forward
direction \cite{CMTV, VM04}. But in accordance with proved analyticity of
$\fsct$ in \lee DR are valid also for some range of negative $t$.

The necessary unitary constrains on partial amplitudes were derived in
previous section. From inequality (\ref{44}) it follows directly that
\be\label{45}
0 \leq \tals \leq 1.
\ee
In accordance with formula (\ref{39}) we have:
\be\label{46}
\casct = 2\,\sum\limits_0^{\infty}\,(2\,l + 1)\,\tals\,\plct.
\ee
From condition (\ref{45}) and the Legendre polynomial properties it follows
directly that
\be\label{47}
\left.\,\frac{d^n\,\casct}{{(d\,\cos\,\vartheta)}^n}\,\right|
_{\cos\,\vartheta = 1} \geq 0.
\ee
The last condition plays a principal role in Martin method of enlarging \leep

One point has to be mentioned. In the extension of Lehmann ellipse, Martin
used the results of Bros, Epstein and Glaser \cite{BEG}. Let us recall
that in this paper analyticity  in $\cos\,\vartheta$ were proved for
nonphysical $s$.
Here we derived analyticity only for physical values of $s$.
Nevertheless following to Sommer \cite{Som}, Bessis and Glaser \cite{Bes}
we can use for this purpose analyticity in Lehmann ellipses only. 
Thus we have no problem with the proof of analyticity of $\fsct$ in \mae 
and so \fmb is valid in space-space NC QFT.

As it was first found by Martin \cite{Mar63a} upper bound (\ref{stf})
in fact contains an additional factor $\ses/\sts$. The strongest bound of
such a kind was obtained in \cite{SR}. Namely,
\be\label{48}
\sts \leq \frac{\pi}{m^2_{\pi}}\,\frac{\ses}{\sts}\,
\ln^2\,\frac s {{(\ln\,s)}^{\frac 3 2}}.
\ee
In \nc case we have also the additional factor, but now
\be\label{49}
\sts \leq \frac{\pi}{m^2_{\pi}}\,\frac{\tses}{\sts}\,
\ln^2\,\frac s {{(\ln\,s)}^{\frac 3 2}}.
\ee
Thus in general  \nc \fmb may have an additional factor,
which is less than unity.

In conclusion let us point out that inequality (\ref{sir}) can be also
proved by standard way. Bounds at $t < 0$ can be obtained directly only
for considered class of scattering amplitudes.

\vskip .5cm 
{\Large\bf Acknowledgments}

We thank M. Chaichian and A. Tureanu for collaboration 
at an early stage of this work.

\section{ Appendix: Derivation of Froissart-Martin Bound }

We start with very weak upper bound on $\castz, \; t_0 = 4\, m^2 - \eps$:
$$
\castz < \exp {(s^n)}, \qquad s \to \infty.          \eqno (A.1)
$$
It was shown by Logunov, Nguyen van Hieu and Todorov \cite{LNT} that this
condition is a sufficient one for derivation of polynomial boundedness of
elastic scattering amplitudes at physical energies.

Indeed, at $s \to \infty$ and $l/{\sqrt s} \to \infty$
$$
P_l\,(x_0) \cong \frac {e^{\gamma}}{\sqrt {2\,\pi\,\gamma}}, \quad
\gamma \equiv 2l\,\sqrt {\frac {t_0} s}, \quad x_0 = 1 + \frac {2t_0} s
                                                            \eqno (A.2)
$$
(see \cite{Roy, VM}).

Let us recall that in accordance with analyticity of $\cast$ in \mae the
series (\ref{46}) converges at $t = t_0$. All terms in this series are
positive according to condition (\ref{45}) since  $P_l\,(x) > 1$ if $x >
1$. Owing to (A.2) bound (A.1) can be satisfied only if
$$
\tilde a_{l' + L} < \exp {\left(- 2\,l'\,\sqrt {\frac {t_0} s}\right)},
\eqno (A.3) 
$$
where $L \sim s^{n + 1/2}$.

Using constrains (\ref{45}) on $\tals$ for $l \leq L$ and inequality
(A.3) for $l \geq L$, it is easy to estimate
$$
\casz = 2\,\sum\limits_0^{\infty}\,(2\,l + 1)\,\tals       \eqno (A.4)
$$
and obtain polynomial boundedness of $\casz$.

In accordance with Jin and Martin result \cite{JM} polynomial boundedness
of $\casz$ leads to polynomial boundedness of $\castz$ as number of
subtractions in DR coincides at $t = 0$ and $t = t_0$ (if this number is
even). Polynomial boundedness of $\castz$  leads to the new constrains on
$\tals$.

Precisely, if $\castz < s^k$ then inequality (A.3) is satisfied if
$$
L = \frac{k - 1/2}{2}\,\sqrt{\frac{s}{t_0}}\,\ln\,s.        \eqno (A.5)
$$
Repeating the above mentioned calculations we obtain Froissart-Martin
bound, but with an unknown constant instead of $\pi/{m_{\pi}^2}$.
Following this way and using now $k = 2$, we come to inequality
(\ref{stlm}) with additional factor $9/4$.

To obtain the strongest upper bound first notice that condition
$$
\intsz \frac {\castz\,\d\,s}{s^3} < \infty          \eqno (A.6)
$$
implies that
$$
\castz < \frac {s^2}{\ln\,s}, \qquad  s \to \infty.         \eqno (A.7)
$$
Taking into account that $\max\,\casz$ is a growing function of $\castz$,
we obtain the desired bound if we find the set of $\tals$ that realize
$\max\,\casz$ at given value of $\castz$. Let us show that this set is:
$$
\tilde a_l\,(s) = \left \{\begin{array}{ll}
1 & l \leq L \\
\eta \leq 1 & l = L + 1  \\
0 & l \geq L + 2
\end{array}\right.                                       \eqno (A.8)
$$
This result follows from the property of the Legendre polynomials:
$P_{l_2}\,(x) > P_{l_1}\,(x)$, if $x > 1$ and $l_2 > l_1$.

Let us prove that any set of $\tals$ different from the set (A.8) can not
realize $\max\,\casz$. Really in any other set there always exist two
partial amplitudes $\tilde a_{l_1}\,(s)$ and $\tilde a_{l_2}\,(s), \; l_2
> l_1$, such that $\tilde a_{l_1}\,(s) < 1$ and $\tilde a_{l_2}\,(s) > 0$.
Let us replace $\tilde a_{l_1}\,(s)$ by $\tilde a_{l_1}\,(s) +
\Delta_1$ and $\tilde a_{l_2}\,(s)$ by $\tilde a_{l_2}\,(s)- \Delta_2,\:
\Delta_i > 0$ in such a way that $\castz$ remains unchanged, i.e.
$$
(2\,l_1 + 1)\,\Delta_1 - (2\,l_2 + 1)\,\Delta_2 \, \frac {P_{l_2}\,(x_0)}
{P_{l_1}\,(x_0)} = 0.
$$
It is evident that
$$
\Delta\,\casz = (2\,l_1 + 1)\,\Delta_1 - (2\,l_2 + 1)\,\Delta_2 > 0.
$$
In order to find $L$ we note that the contribution from the
partial amplitude with $l = L + 1$ can be neglected, and because of the
known recursion formula
$$
(2\,l + 1)\,\pl = \plppo - \plpmo
$$
the Legendre polynomial series of $\castz$ can be summed up. As a result
we have the following equation on $L$:
$$
A\,(s, t) =  P_{L+1}'\,(x) + P_{L}'\,(x) \cong 2\, P_{L}'\,(x) \eqno (A.9)
$$
From (A.2) it follows that
$$
P'_l\,\left (1+\frac {2t_0} s\right) \cong \frac {e^\gamma \,\sqrt
\gamma}{4\,\sqrt{2\pi}}\:\frac s {t_0},  \qquad \gamma = 2 L\, \sqrt{\frac
{t_0} s}.                                              \eqno (A.10)
$$
Thus in accordance with (A.9) and (A.7)
$$
e^{\gamma}\,\sqrt {\gamma} \cong \frac{s}{\ln\,s}.      \eqno (A.11)
$$
This equation is easily solved by the method of successive approximations.
As a result we obtain that at $s \to \infty$
$$
\gamma \cong \ln\,\frac s {{(\ln\,s)}^{\frac 3 2}}.       \eqno (A.12)
$$
(For the details see \cite{VM}).

According to the optical theorem we have at $ t = 0$:
$$
\sigma_{tot}^{max}\,(s) \cong
\frac {32\,\pi}{s} \sum \limits _{l=0}^L 2\,l \cong
\frac {16\,\pi}{s}\,L^2 = \frac {4\,\pi} {t_0}\, {\gamma}^2. \eqno (A.13)
$$
Taking into account (A.12) we see that equality (A.13) implies that
desired  inequality (\ref{stlms}) is fulfilled.

Let us point out that maximum of $\sigma_{tot}\,(s)$ is reached if $\ses =
\sts$, see eqs. (\ref{42}) - (\ref{44}).

\end{document}